# First comparison of composite 0.52 – 55 keV ENA spectra observed by IBEX and Cassini/INCA with simulated ENAs inferred by proton hybrid simulations downstream of the termination shock.

Matina Gkioulidou[1], Merav Opher[2,3], M. Kornbleuth[2], K. Dialynas[4], J. Giacalone[5], J. D. Richardson[6], G. P. Zank[7], S. A. Fuselier[8,9], D. G. Mitchell[1], S. M. Krimigis[1,4], E. Roussos[10], I. Baliukin[11,12,13]

[1]Applied Physics Laboratory, Johns Hopkins University, Laurel, MD 20723, USA

[2]Astronomy Department, Boston University, Boston, MA 02215, USA

[3]Radcliffe Institute for Advanced Study at Harvard University, Cambridge, MA, USA

[4]Office of Space Research and Technology, Academy of Athens, 10679 Athens, Greece

[5]Lunar & Planetary Laboratory, University of Arizona, Tucson, AZ 85721, USA

[6]Kavli Institute for Astrophysics and Space Research and Department of Physics, Massachusetts Institute of Technology, Cambridge, MA, USA

[7]Department of Space Science, The University of Alabama in Huntsville 4 Huntsville, AL 35805, USA

[8]Southwest Research Institute, San Antonio, TX 78228, USA

[9]University of Texas at San Antonio, San Antonio, TX 78249, USA

[10]Max Planck Institute for Solar System Research, Justus-von-Liebig-Weg 3, D-37077 Goettingen, Germany

[11]Space Research Institute of Russian Academy of Sciences, Profsoyuznaya Str. 84/32, Moscow, 117997, Russia

[12]Moscow Center for Fundamental and Applied Mathematics, Lomonosov Moscow State University, GSP-1, Leninskie Gory, Moscow, 119991, Russia

[13]HSE University, Moscow, Russia

Corresponding author: M. Gkioulidou, Johns Hopkins University-Applied Physics Laboratory, 11100 Johns Hopkins Rd., Laurel, MD 20723-6099, USA (matina.gkioulidou@jhuapl.edu)

**Abstract**

We present a first comparison of Energetic Neutral Atom (ENA) heliosheath measurements, remotely sensed by the Interstellar Boundary Explorer (IBEX) mission and the Ion and Neutral Camera (INCA) on the Cassini mission, with modeled ENA inferred from interstellar pickup protons that have been accelerated at the termination shock, using hybrid simulations. The observed ENA intensities are an average value over the time period from 2009 to the end of 2012, along the Voyager 2 trajectory. The hybrid simulations parameters for the solar wind, interstellar pickup ions (PUIs), and magnetic field upstream of the termination shock, where Voyager 2 crossed, are based on observations. We report an energy dependent discrepancy between observed and simulated ENA fluxes, with the observed ENA fluxes, being consistently higher than the simulated ones, and discuss possible causes of this discrepancy.

1. **Introduction**

As the solar system and its surrounding heliosphere move through the local interstellar medium (LISM), interstellar neutral (ISN) atoms, mostly atomic Hydrogen with densities of > 0.12 /cm$^3$ (Dialynas et al. (2019); Swaczyna et al. (2020)), enter the heliosphere and undergo charge-exchange collisions with the continuously flowing solar wind (SW) protons. During this process, SW protons gain an electron and become neutral hydrogen, still flowing outward at the SW velocity. Newly created ions from the ISN population are advected outward with the SW under the force of the **V**x**B** electric field, typically as a ring-beam distribution that is isotropized by scattering from self-excited and pre-existing magnetic fluctuations, forming a population that is commonly known as pickup ions (PUIs). Recent observations from the New Horizons spacecraft at ~38 AU (McComas et al., 2017) showed that PUIs are heated in the frame of the SW with increasing distance (McComas et al., 2021), before reaching the termination shock (TS). At the shock, they are further heated, with a fraction of their distribution being reflected off the shock surface and undergoing additional heating, (Zank et al 1996, 2010).

The two Voyager spacecraft (V1 and V2) reached the TS in 2004 and 2007 at distances of ~94 (Decker et al., 2005; Stone et al., 2005) and ~84 Astronomical Units (AU) (Decker et al., 2008), respectively. Voyager observations revealed that the TS was mediated by PUIs (in agreement with theoretical prediction of Zank et al. 1996, 2010), where roughly 80% of the solar wind flow energy was transferred to PUIs in the heliosheath (HS), the region between the TS and the heliopause (i.e. the interface between our heliosphere and the very local interstellar medium (VLISM)), including a substantial part (>15%) that went into >28 keV protons (Richardson et al. 2008; Decker et al. 2008). The heliopause was observed at distances of ~122 AU (Krimigis et al. 2013; Stone et al. 2013; Burlaga et al. 2013; Gurnett et al. 2013) and ~119 AU (Krimigis et al.

2019; Stone et al. 2019; Richardson et al. 2019; Gurnett & Kurth 2019; Burlaga et al 2019) from V1 and V2, respectively.

The heated PUIs that populate the HS, charge-exchange with the interstellar neutrals, and are measured remotely by the Interstellar Boundary Explorer (IBEX; 0.01-6 keV) and Cassini/Ion and Neutral Camera (INCA; 5.2-55 keV), providing Energetic Neutral Atom (ENA) full-sky maps (McComas et al. 2009; Krimigis et al. 2009). These images showed the existence of a bright and narrow ribbon (Schwadron et al. 2009) of ENA emissions that is measured by IBEX-Hi and is thought to lie beyond the HP, formed through a secondary ENA process (e.g. Heerikhuisen et al. 2010; McComas et al. 2017) and the Globally Distributed Flux (GDF; Schwadron et al. 2011), a "background" ENA flux, with the IBEX ribbon removed, that is largely produced in the HS (Zank et al., 2010, Dayeh et al. 2011; Zirnstein et al. 2020; McComas et al. 2020). Spectral agreements between Cassini/INCA ENA measurements and the Voyager ion measurements suggest that the source of ENA emissions at energies above 5.2 keV are most likely produced by charge exchange interactions inside the HS (Dialynas et al. 2013; 2017). Therefore, understanding the PUI distribution downstream of the TS is essential to study the pressure balance and acceleration mechanisms inside the HS (Dialynas et al. 2019; 2020). This understanding is needed to determine the emission of energetic neutral atoms (ENAs) from the HS, because these ENAs are used to remotely sense the boundaries of our heliosphere and its interaction with the VLISM.

This letter provides an unprecedented comparison between observed ENAs in the HS (Section 3), combining IBEX-Hi measurements of energies from 0.52 to 6 keV with higher energy ENAs from Cassini/INCA from 5.2 to 55 keV, and modeled ENA spectra inferred from hybrid simulations of interstellar pickup protons accelerated at the TS using realistic parameters (Section 2). Our comparison shows that the observed ENA fluxes are higher than the modeled ones throughout the 0.52-55 keV energy range and that the identified discrepancy is energy dependent. We conclude (Section 4) with a short discussion concerning possible causes of that discrepancy which will hopefully guide future analyses.

## 2. Simulated PUI distribution

We use the hybrid model by Giacalone et al. (2021) to produce interstellar pick up protons downstream the TS, which have been reflected and some further accelerated at the TS (Zank et al., 1996). The hybrid simulation is two-dimensional, however, vectors, such as average proton velocity, magnetic field, electric field, etc., are not confined to the simulation plane and they point in three directions. The initial magnetic field and bulk plasma velocity consists of an average component and a turbulent component. The simulation is a self-consistent, kinetic treatment of SW protons and pickup protons, and massless, charge-neutralizing fluid SW electrons. Giacalone et al. (2021) determined the intensity of accelerated interstellar PUIs at three different locations: the Voyager 2 (V2) crossing, the flank, and the tail. For the purposes of this study, we use the results at the Voyager 2 crossing location, where simulation parameters are based on V2 observations just before the TS crossing (see second column of Table 1 in Giacalone et al. (2021)).

Figure 1 shows energy spectra from the hybrid simulation for two proton populations, the SW (green dotted lines) and PUI (purple dotted lines), in the shock frame. These spectra are averaged over nearly the entire downstream region of the simulation. The simulations in Figure 1 show high energy tails that are formed for both the SW (starting at ~ 3 keV) and PUI (starting at ~ 5 keV) distributions, with the intensity of accelerated SW protons being significantly lower than that of the PUIs. The simulated PUI intensity decreases significantly above ~50 keV, due to artificial losses arising from the small simulation domain and run time (Giacalone et al. 2021). As a result, there is a substantial mismatch between the simulated spectra and the ion measurements from the Low Energy Charged Particle (LECP) detector on Voyager 2 (e.g. Decker et al. 2008; Dialynas et al. 2019). It should be noted, however, that the simulated PUI tail intensities in Giacalone et al. (2021) right before the abrupt decrease, i.e., up to ~ 80 keV, agree well with the corresponding LECP intensities as measured downstream of the TS.

We then use these PUI intensities produced by the hybrid model downstream the TS, to infer ENA intensities throughout the HS. We use the ENA model described in Kornbleuth et al. (2021a), where, bulk plasma velocity streamlines are extracted from the BU MHD model simulations, described in section 3, in order to simulate ions crossing the TS and transiting through the heliosphere until they charge exchange with interstellar neutrals and become ENAs.

In the ENA model, similarly to the methodology followed in Zank et al., (2010) and Zirnstein et al., (2017), PUIs and SW protons are treated as one fluid, and this requires partitioning of the plasma energy among three different ion populations, namely, i) thermal SW protons, ii) PUIs created in the supersonic SW and being adiabatically transmitted across the TS, and iii) PUIs being reflected at the TS until they have sufficient energy to overcome the cross-shock potential barrier (Zank et al., 1996). We partition the total thermal energy of the plasma via:

$$T_p = \left(\frac{n_{SW}}{n_p}\Gamma_{SW} + \frac{n_{tr}}{n_p}\Gamma_{tr} + \frac{n_{energ}}{n_p}\Gamma_{energ}\right) T_p \qquad (1)$$

where $n_p$ and $T_p$ are the density and temperature of the plasma, $n_i$ is the density of the respective ion population, and $\Gamma_i$ is the temperature fraction, $T_i/T_p$, where $T_i$ is the temperature of the respective ion population. We model downstream plasma conditions in the Voyager 2 direction by fitting the ion spectra produced by the hybrid simulation with a Maxwellian distribution in the case of cold SW protons, with $n_{SW} = 0.74 * n_p$ and thermal energy $E_{SW} = 0.07 * E_p$, and kappa distributions $\left(f(v) = \frac{n_i}{W^3 \pi^{\frac{3}{2}}} \frac{\Gamma(\kappa+1)}{\Gamma\left(\kappa-\frac{1}{2}\right)} \frac{1}{\kappa^{\frac{3}{2}}} \left(1 + \frac{v^2}{\kappa W^2}\right)^{-\kappa-1}; where\ W = \left[\frac{2\left(\kappa-\frac{3}{2}\right)E_i}{\kappa\ m\ n_i}\right]^{1/2}\right)$ for the transmitted (energized) PUIs with parameter $\varkappa = 10\ (1.57)$, density $n_{PUI_{trans}(PUI_{energ})} = 0.23\ (0.03) * n_p$, and energy $E_{PUI_{trans}(PUI_{energ})} = 0.5\ (0.43) * E_p$, where $n_p$ is the plasma density and $E_p$ is the thermal energy of the plasma extracted from the BU MHD model of Kornbleuth et al. (2021b). These fits are depicted with dashed lines in Figure 1 (green for SW; blue for transmitted PUIs; red for the reflected and energized PUIs). The purple solid line shows the total PUI population spectra, combining the fits of the transmitted and energized populations. By fitting these distribution functions, we are able to determine density and temperature ratios for each population downstream of the TS, which is used in ENA modeling, as described. The use of the hybrid simulation from Giacalone et al. (2021) assures a reasonable agreement of the simulated fluxes with the measured LECP ions (within a limited energy range of ~28 to <80 keV), while treating the thermal solar wind protons and PUIs separately, allows us to constrain high energy ion properties downstream of the TS. In our simulations we assume quasi-neutrality, i.e. $n_e = n_p = n_{Sw} + n_{PUI_{trans}} + n_{PUI_{energ}}$, and that electrons and solar wind protons have the same temperature.

3. **Comparison of modeled and observed ENA spectra**

Figure 2a shows the combined IBEX-Hi (0.52-6 keV) and INCA (5.2-55 keV) ENA spectra, with orange and red symbols, respectively. The ENA observations, are taken from the beginning of 2009 to the end of 2012 (adopted from Dialynas et al. (2020)). The 0.52-6 keV ENAs are measured from IBEX-Hi (Ram-Only direction; data release 16; McComas et al. 2020), that has a roughly circular instantaneous field of view that is 6.5° (FWHM) wide, viewed perpendicular to the spacecraft spin axis (Funsten et al. 2009; see also Table 3; Qualified Triple-Coincidences with background removed as in McComas et al. 2014). IBEX-Hi data are corrected for the survival probability (Bzowski 2008; ionization rate model shown in Sokol et al. 2020; radiation pressure model from Kowalska-Leszczyńska et al. 2020) and the motion of the IBEX spacecraft relative to the Sun (Compton & Getting 1935). The 5.2-55 keV ENAs are measured from Cassini/INCA (Krimigis et al. 2009), part of the Magnetospheric Imaging Instrument (MIMI; Krimigis et al. 2004). INCA utilizes a broad Field Of View (FOV) of 90° in the nominal Cassini roll direction and 120° in the direction perpendicular to the spacecraft roll plane, and analyzes separately the composition (H and O groups), velocity, and direction of the incident ENAs, based on the time-of-flight (TOF) technique. Pixels pointing towards the solar disk and/or Saturn's magnetosphere have been excluded (see Methods in Dialynas et al. 2017), because of possible background contamination. INCA data were successfully cross-calibrated with data from the Low-Energy Magnetospheric Measurement System (LEMMS) and the Charge-Energy Measurement System (CHEMS) that complemented the MIMI suite.

These observed ENA fluxes are directly contrasted with simulated ENA fluxes produced by the same ENA model used in Kornbleuth et al. (2021a). More specifically, we fit the PUI densities and temperatures just downstream of the TS from the BU MHD simulations with the hybrid results, as described in Section 2, based on ratios relative to plasma density and temperature. Then those PUIs are transported further into the HS, following bulk plasma velocity streamlines, extracted from the BU MHD model simulations, until they charge exchange with interstellar neutrals and become ENAs.

At the outer boundary of the BU MHD simulation (1500 AU from the sun), we assume an interstellar proton density of $n_{p,LISM} = 0.04 \, cm^{-3}$ and interstellar neutral hydrogen density of $n_{H,LISM} = 0.14 \, cm^{-3}$ (lower than those reported by Swaczyna et al. (2020), i.e. ~0.19 cm$^{-3}$), based on the study of Izmodenov & Alexashov (2015) who used these parameters to best fit the TS distance with respect to Voyager measurements. The neutral and ionized populations in the interstellar medium are assumed to have the same bulk velocity and temperature at the outer boundary in the pristine ISM, given by $v_{ISM}$ = 26.4 km/s (longitude = 75.4°, latitude = -5.2° in ecliptic J2000 coordinate system) and $T_{ISM}$ = 6530 K, respectively. Based on the work of Izmodenov & Alexashov (2020) the BU MHD simulation uses 22-year averaged solar cycle conditions from the years 1995 to 2017, with heliolatitudinal variations of the solar wind speed and density based on interplanetary scintillation data (Tokumaru et al. 2012) and SOHO/SWAM full-sky maps of backscattered Lyman-alpha intensities (QuéMerais et al. 2006; Lallement et al. 2010; Katushkina et al. 2013, 2019).

The proton distributions predicted by the hybrid model (SW, and transmitted and energized PUIs) are transported within the HS following the MHD streamlines, even though that assumption might not be exactly accurate for the most energetic ones, and they charge exchange with interstellar neutral hydrogen producing ENAs. The ENA flux along a radial line-of-sight (LOS) is given by:

$$J(E,\theta,\varphi) = \int_{r_{observer}}^{\infty} \frac{2E}{m_p^2} f_p\left(n_i(\boldsymbol{r}'(s)), T_i(\boldsymbol{r}'(s)), v_{plasma}(\boldsymbol{r}'(s))\right) * n_H(\boldsymbol{r}'(s)) \sigma_{ex}(E) S(E) \, d\boldsymbol{r}'(s) \quad (2)$$

where $\boldsymbol{r}'$ is the vector along a particular LOS as a function of $\theta$ and $\varphi$ and $s$ is the distance along the vector, $m_p$ is the mass of a proton and $f_p$ is the phase space velocity distribution, which is treated as a Maxwellian for SW protons, and as a kappa for PUIs, as described in Section 2. The TS of the MHD model is at $82 \pm 4$ AU, and the heliopause at $126 \pm 1$ AU. The integral only captures 44 AU in the HS along the V2 LOS, which is larger than the width of the HS as measured by V2, which crossed the TS at ~84, and the heliopause at ~119 AU. The velocity of the protons in the frame of the plasma is given by $v_{plasma} = |\mathbf{v}_p - \mathbf{v}_i|$, where $\mathbf{v}_p$ and $\mathbf{v}_i$ are the velocities of the bulk plasma and the parent proton, respectively. For the density and temperature of the given proton population, we use $n_i$ and $T_i$, respectively, as described in the previous section.

$n_H\left(\mathbf{r}'(s)\right)$ is the neutral H density along the line of sight, $\sigma_{ex}(E)$ is the charge-exchange cross section from Lindsay & Stebbings (2005) and $S(E)$ is the survival probability, which represent the likelihood that an ENA created in the HS will charge exchange prior to being observed at 1 AU. The survival probability of an ENA is calculated along the radial LOS:

$$S(E) = \exp\left(-\int_{r_{source}}^{r_{observer}} \frac{\sigma(v_{rel})v_{rel}n_p}{v_{ENA}} dr\right) \quad (3)$$

where $dr$ is the radial element over which we are integrating, $v_{ENA}$ is the speed of the ENA, and $v_{rel}$ is the relative velocity between the ENA and the bulk plasma given by Pauls et al., (1995),

$$v_{rel} = v_{th,p}\left[\frac{e^{-\omega^2}}{\sqrt{\pi}} + \left(\omega + \frac{1}{2\omega}\right)\text{erf}(\omega)\right], \omega = \frac{1}{v_{th,p}}\left|\mathbf{v}_{ENA} - \mathbf{u}_p\right| \quad (4)$$

where $\mathbf{v}_{ENA}$ is the velocity of the ENA, $\mathbf{u}_p$ is the bulk averaged plasma velocity, and $v_{th,p}$ is the thermal speed of the plasma. In calculating the survival probability, we assume the observer to be located at 100 AU, since the ENA observations are corrected for survivability from 100 AU to the inner solar system.

The modeled and observed ENA fluxes in Figure 2a show a clear energy dependent discrepancy throughout the 0.52-55 keV energy range (blue shaded area), with the observations being consistently higher. The ratios of the observed ENA fluxes over the modeled ones, as a function of energy, are shown in Figure 2b (and embedded table). Those ratios increase with energy for ENAs with energies from 0.71 and higher, peaking at about 8.38 keV, and it start decreasing with energy beyond this point. The modeled ENA spectra also exhibit different slopes than the measured ENA spectra throughout the 0.52-55 keV energy range. The resulting spectra from the model are softer (harder) than the observed ones in the 0.52 – 8 keV energy range (beyond ~ 8 keV), i.e. the ratio between the measurements and the model within that energy range is increasing (decreasing). The overall best comparison between the model and the measurements occurs for >18 keV, which is not surprising when considering that i) the PUIs from the hybrid simulation that we use (Giacalone et al. 2021) agree well with the ~28 keV measurements from LECP in V2, and ii) the conversions of the in-situ ~28 to 540 keV LECP ions to ENAs (Dialynas et al. 2020) result in ENA spectra that retain similar power law slopes as the measured 5.2–55 keV ENA spectra from Cassini/INCA (Krimigis et al. 2009; Dialynas et al. 2020).

## 4. Discussion and Conclusions

We have compared observed ENAs from IBEX-Hi (0.52 – 6 keV) and Cassini/INCA (5.2 – 55 keV), with modeled ENA spectra inferred from interstellar pickup protons that have been accelerated at the TS using hybrid simulations with realistic parameters. Our comparison shows that there is an energy dependent discrepancy between the modeled and observed ENAs throughout the 0.52-55 keV energy, with the observed ENAs being consistently higher than the modeled ones. The disagreement with observations potentially indicates that an additional physical process, energizing further the PUIs, has yet to be identified. Such a process would also require a source of energy. In the hybrid simulation, energy is conserved, and it is the upstream solar wind ram energy that is accelerating the ions. Even though the hybrid model is local, it captures the main components of turbulence (that is <0.01 of the ram pressure), so the distribution of energy across the TS is captured correctly. How the redistribution of energies further downstream in the HS (including solar wind ram pressure, thermal pressure etc), as well as any other source of energy, yet to be identified, could further accelerate the PUIs, is still an open question. Nonetheless, it should be noted that for ENAs of energies < 1keV, an alternative possible reason for the data-model discrepancy could be the source of these ENAs not being the HS but the VLISM, as discussed in Fuselier et al. (2021).

Using test particle simulations, Zirnstein et al. (2021) showed that although a moderate level of turbulence with a power ratio of $\left(\frac{\Delta B}{B_o}\right)^2 \cong 0.01$ (based on V2/MAG measurements from Burlaga et al. (2008) at the TS), was sufficient to reproduce a superthermal PUI tail downstream of the TS, the derived intensities were lower than those observed by IBEX-Hi. In fact, to produce a proton distribution consistent with IBEX observations, a ten times larger turbulence power ratio must be applied at the shock foot. The authors speculated that this enhanced level of turbulence in the PUI foot may be related to shock self-reformation (Lembege et al. 2004), shock front ripples (Umeda et al. 2014) originating from instabilities induced by ion temperature anisotropies (Winske & Quest 1988), cross field currents (Lembege & Savoini 1992), or interplanetary SW turbulence transported through the TS (Giacalone et al. 2021; Zank et al., 2021).

In this study, we have used a wider range of ENA energies, including also higher energy ENAs from the Cassini/INCA instrument (~5.2-55 keV), and we also used a self-consistent hybrid simulation to produce the PUI population in the HS. Similarly to Zirnstein et al. (2021), we find that even self-consistent hybrid simulations do not accelerate PUIs enough at the TS, in order to agree with the observed spectra. Furthermore, we find that the data-model discrepancy exhibits a clear energy dependence, where the modeled ENA spectra do not capture the observed spectral slopes. Thus, we infer that an additional acceleration region for these low energy PUIs is most likely the HS.

Models have showed that processes like turbulence or magnetic reconnection (Drake et al. 2010; Opher et al. 2011) may occur throughout the HS, as opposed to only close to TS. Zhao et al., (2019) analyzed Voyager 2 data, and found evidence of magnetic flux ropes/islands, accompanied by energetic protons, suggesting reconnection within the HS. The source of that HS turbulence has been investigated in several recent studies. Zieger et al. (2020) found that dispersive fast magnetosonic waves can be responsible for energy dissipation downstream of the TS. Zank et al. (2018) explored the transmission of nearly incompressible turbulence across the TS. Compressive turbulence within the HS may also further heat the PUIs, after crossing the TS (Fisk & Gloeckler, 2017). Zank et al. (2021) studied the interaction and transmission of quasi-2D turbulence, through a collisionless perpendicular shock wave in the large beta regime, and found that the downstream spectral amplitude is increased significantly. Simulations also demonstrate that a Rayleigh-Taylor like instability (Opher et al. 2021), which mixes the HS and LISM plasmas, can form turbulent heliospheric jets with scales of order 100 AU and a turnover timescale of years (Opher et al. 2015). How far downstream the turbulence ensues, whether the above processes could further accelerate the PUIs in sufficient numbers, depending on the source of energy, and how these accelerated PUIs get transported further into the HS should be investigated in future studies.

Although exploring each of the processes mentioned above, which could potentially act to further accelerate PUIs within the HS, lies beyond the scope of the present letter, we anticipate that the reported discrepancy will guide relevant future modeling efforts. Such investigations will be especially crucial when data from the upcoming Interstellar Mapping and Acceleration Probe mission (IMAP; McComas et al. 2018) will become available, covering ENAs of energies from 0.005 – 300 keV taken from instruments on the same platform in L1 orbit, offering major advancements compared to previous observations, such as significant increase in the collection power of all the ENA cameras onboard.

This work was supported by the NASA Drive Science Centers initiative, under contract 80NSSC20K0603-18-DRIVE18_2-0029 in BU SHIELD DRIVE Science Center (http://sites.bu.edu/shield-drive/). The work at the Office for Space Research and Technology was supported by subcontract to JHU/APL (NASA contracts NAS5 97271, NNX07AJ69G, and NNN06AA01C). MG was also supported by NASA IMAP contract 80MSFC20D0004. MO was also supported by Fellowship Program, Radcliffe Institute for Advanced Study at Harvard University.

**References**

Burlaga, L. F., Ness, N. F., Acuña, M. H., et al. 2008, Natur, 454, 75

Burlaga, L. F., Ness, N. F., & Stone, E. C. 2013, Sci, 341, 147

Burlaga, L. F., Ness, N. F., Berdichevsky, D. B., et al. 2019, NatAs, 3, 1007

Bzowski, M. 2008, A&A, 488, 1057

Compton, A. H., & Getting, I. A. 1935, PhRv, 47, 818

Dayeh, M. A., McComas, D. J., Livadiotis, G., et al. 2011, ApJ, 734, 29

Decker, R. B., Krimigis, S. M., Roelof, E. C., et al. 2005, Sci, 309, 2020

Decker, R. B., Krimigis, S. M., Roelof, E. C., et al. 2008, Natur, 454, 67

Dialynas, K., Krimigis, S. M., Mitchell, D. G., Roelof, E. C., & Decker, D. B. 2013, ApJ, 778, 13

Dialynas, K., Krimigis, S. M., Mitchell, D. G., Decker, R. B., & Roelof, E. C. 2017, NatAs, 1, 0115

Dialynas, K., Krimigis, S. M., Decker, R. B., & Mitchell, D. G. 2019, GeoRL, 46, 7911

Dialynas, K., Galli, A., Dayeh, M., et al. 2020, ApJL, 905, L24

Dialynas, K., Krimigis, S. M., Decker, R. B., & Hill, M. E. 2021, ApJ, 917, 42

Drake, J. F., Opher, M., Swisdak, M., & Chamoun, J. N. 2010, ApJ, 709, 963

Funsten, H. O., Allegrini, F., Bochsler, P., et al. 2009, SSRv, 146, 75

Giacalone, J., Nakanotani, M., Zank, G. P., et al. 2021, ApJ, 911, 27

Gurnett, D. A., Kurth, W. S., Burlaga, L. F., & Ness, N. F. 2013, Sci, 341, 1489

Gurnett, D. A., & Kurth, W. S. 2019, NatAs, 3, 1024

Heerikhuisen, J., Pogorelov, N. V., Zank, G. P., et al. 2010, ApJ, 708, L126

Izmodenov, V. V., & Alexashov, D. B. 2015, ApJS, 220, 32

Izmodenov, V. V., & Alexashov, D. B. 2020, A&A, 633, L12

Katushkina, O., Izmodenov, V., Koutroumpa, D., et al. 2019, SoPh, 294, 17

Katushkina, O. A., Izmodenov, V. V., Quemerais, E., et al. 2013, JGRA, 118, 2800

Kornbleuth, M., Opher, M., Baliukin, I., et al. 2021a, ApJ, 921, 164

Kornbleuth, M., Opher, M., Baliukin, I., et al. 2021b, ApJ, 923, 13

Kowalska-Leszczyńska, I., Bzowski, M., Kubiak, M. A., & Sokół, J. M. 2020, ApJS, 247, 62

Krimigis, S. M., Mitchell, D. G., Hamilton, D. C., et al. 2004, SSRv, 114, 233

Krimigis, S. M., Mitchell, D. G., Roelof, E. C., Hsieh, K. C., & McComas, D. J. 2009, Sci, 326, 971

Krimigis, S. M., Decker, R. B., Roelof, E. C., et al. 2013, Sci, 341, 144

Krimigis, S. M., Decker, R. B., Roelof, E. C., et al. 2019, NatAs, 3, 997

Lallement, R., Quémerais, E., Koutroumpa, D., et al. 2010, in AIP Conf. Proc. 1216, Twelfth Int. Solar Wind Conf. (Melville, NY: AIP), 555

Lembege, B., Giacalone, J., Scholer, M., et al. 2004, SSRv, 110, 161

Lembege, B., & Savoini, P. 1992, PhFlB, 4, 3533

McComas, D. J., Allegrini, F., Bochsler, P., et al. 2009, Sci, 326, 959

McComas, D. J., Allegrini, F., Bzowski, M., et al. 2014, ApJS, 213, 20

McComas, D. J., Zirnstein, E. J., Bzowski, M., et al. 2017, ApJS, 233, 8

McComas, D. J., Chodas, M., Bayley, L., et al. 2018, SSRv, 214, 8

McComas, D. J., Bzowski, M., Dayeh, M. A., et al. 2020, ApJS, 248, 26

McComas, D. J., Swaczyna, P., Szalay, J. R., et al. 2021, ApJS, 254, 19

Opher, M., Drake, J. F., Swisdak, M., et al. 2011, ApJ, 734, 71

Opher, M., Fisher, R., García-Berro, E., et al. 2015, ApJL, 800, L7

Opher, M., J. F. Drake, G. Zank, et al., ApJL, in press 2021 (https://doi.org/10.21203/rs.3.rs-198925/v1)

QuéMerais, E., Lallement, R., Ferron, S., et al. 2006, JGRA, 111, A09114

Richardson, J. D., Kasper, J. C., Wang, C., Belcher, J. W., & Lazarus, A. J. 2008, Natur, 454, 63

Richardson, J. D., Belcher, J. W., Garcia-Galindo, P., & Burlaga, L. F. 2019, NatAs, 3, 1019

Schwadron, N. A., Bzowski, M., Crew, G. B., et al. 2009, Sci, 326, 966

Schwadron, N. A., Allegrini, F., Bzowski, M., et al. 2011, ApJ, 731, 56

Sokół, J. M, McComas, D. J., Bzowski, M., & Tokumaru, M. 2020, ApJ, 897, 179

Stone, E. C., Cummings, A. C., McDonald, F. B., et al. 2005, Sci, 309, 2017

Stone, E. C., Cummings, A. C., McDonald, F. B., et al. 2013, Sci, 341, 150

Stone, E. C., Cummings, A. C., Heikkila, B. C., et al. 2019, NatAs, 3, 1013

Swaczyna, P., McComas, D. J., Zirnstein, E. J., et al. 2020, ApJ, 903, 48

Tokumaru, M., Kojima, M., & Fujiki, K. 2012, JGRA, 117, A06108

Umeda, T., Kidani, Y., Matsukiyo, S., & Yamazaki, R. 2014, PhPl, 21, 022102

Winske, D., & Quest, K. B. 1988, JGR, 93, 9681

Zank, G. P., Pauls, H. L., Cairns, I. H., & Webb, G. M., 1996, JGR, 101, 457

Zank, G. P., Heerikhuisen, J., Pogorelov, N. V., et al. 2010, ApJ, 708, 1092

17
Zank, G. P., Adhikari, L., Zhao, L.-L., et al. 2018, ApJ, 869, 23

Zank, G. P., Nakanotani, M., Zhao, L. L., et al. 2021, ApJ, 913, 127

Zhao, L. L., Zank, G. P., Hu, Q., et al. 2019, ApJ, 886, 144

Zieger, B., Opher, M., Tóth, G., & Florinski, V. (2020), JGR, 125, e2020JA028393.

Zirnstein, E. J., Heerikhuisen, J., Zank, G. P., et al. 2017, ApJ, 836, 238

Zirnstein, E. J., Giacalone, J., Kumar, R., et al. 2020, ApJ, 888, 29

Zirnstein E. J., Kumar, R., Bandyopadhyay, R., et al. 2021 ApJL 916, L21

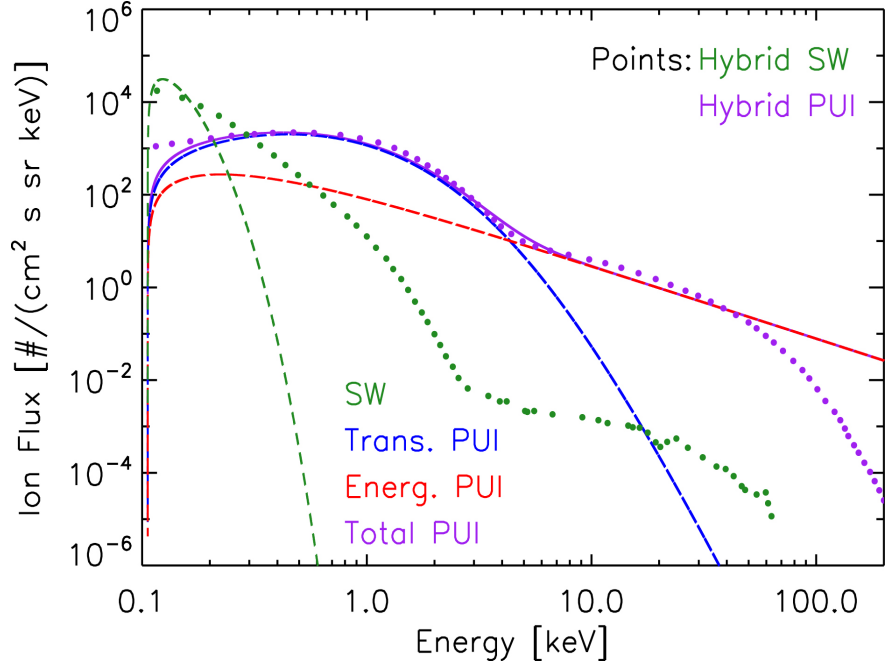

Figure 1: Flux as a function of energy for SW protons (green dotted line) and and PUI (purple dotted line) averaged over nearly the entire downstream region of a hybrid model (Giacalone et al. 2021). Dashed lines represent fits of the hybrid spectra, using downstream plasma conditions in the Voyager 2 direction from the BU MHD model (Kornbleuth et al. 2021b), assuming three different populations: thermal SW protons (green), PUIs created in the supersonic SW, which are i) adiabatically transmitted across the TS (transmitted PUIs in blue), or ii) reflected at the TS until they have sufficient energy to overcome the cross-shock potential (energized PUIs in red). The purple solid line represents the total PUI population, combining the fits of the transmitted and energized populations.

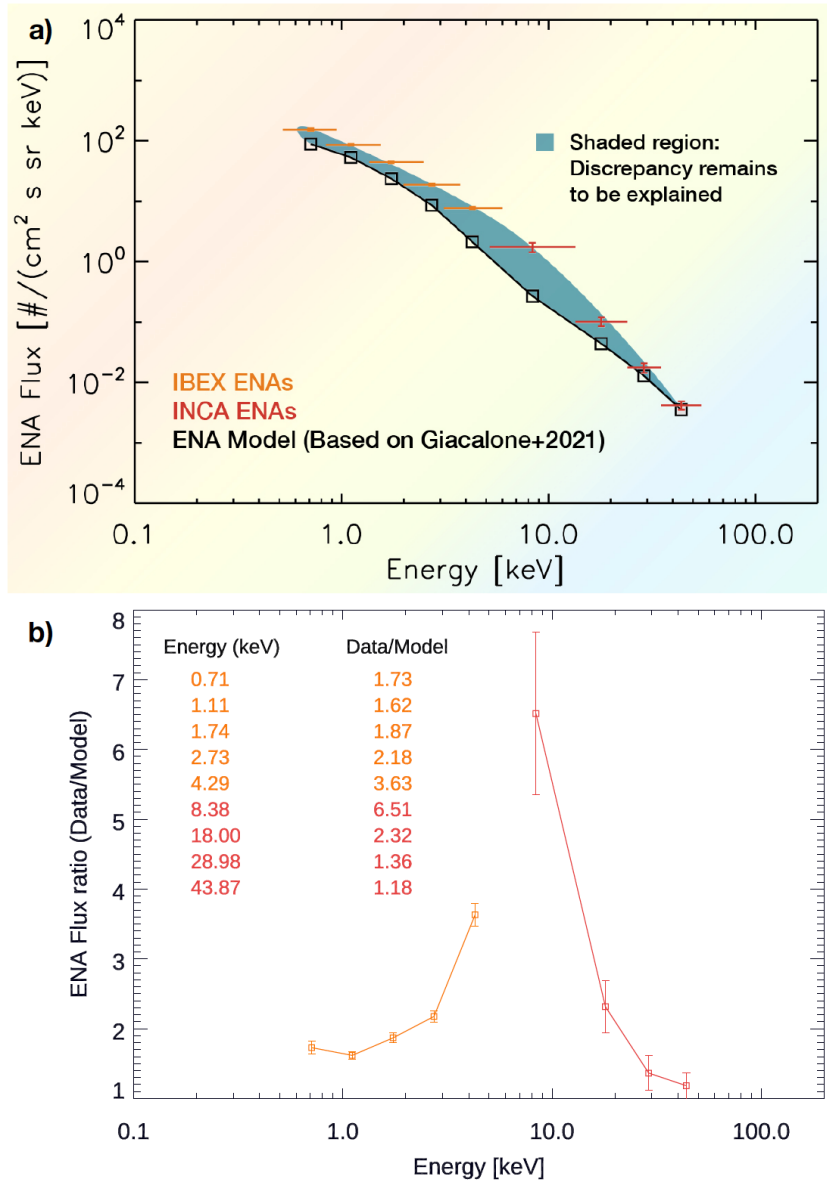

Figure 2: a) Modeled ENA fluxes, as a function of energy, calculated from simulated ion distributions using they hybrid model of Giacalone et al. (2021) (black symbols); ENA fluxes as a function of energy, as observed by IBEX-Hi (0.52 – 6 keV) (orange symbols), and by Cassini/INCA (5.2 – 55 keV) (red symbols). The observed spectra in both cases are averaged within the pixels enclosing the position of Voyager 2, over 9º for IBEX-Hi, avoiding areas in the sky that include the "ribbon", and over 5º for INCA, from the beginning of 2009 to the end of 2012 (left panel of Figure 1 in Dialynas et al., 2020). The horizontal lines represent the energy range of the particularly measurements, while the perpendicular lines represent the measurement uncertainties (mostly smaller than the plot symbol); b) Ratios of observed ENA fluxes over modeled ones, as a function of energy